\documentclass[english,superscriptaddress,amssymb,twocolumn]{revtex4}

\usepackage{bm}
\usepackage{graphicx}
\usepackage{color}
\usepackage{amsmath}

\usepackage[T1]{fontenc}
\usepackage[latin9]{inputenc}
\usepackage{textcomp}
\usepackage{amssymb}
\usepackage{ulem}
\usepackage{extraplaceins}
\makeatother

\usepackage{pdfpages}
\usepackage[english]{babel}

\begin{document}

\title{Observing interferences between past and future quantum states in resonance fluorescence}

\author{P. Campagne-Ibarcq}
\thanks{These two authors contributed equally to this work}

\affiliation{Laboratoire Pierre Aigrain, Ecole Normale Supérieure, CNRS (UMR 8551),
Université P. et M. Curie, Université D. Diderot 24, rue Lhomond,
75231 Paris Cedex 05, France }

\author{L. Bretheau}
\thanks{These two authors contributed equally to this work}

\affiliation{Laboratoire Pierre Aigrain, Ecole Normale Supérieure, CNRS (UMR 8551),
Université P. et M. Curie, Université D. Diderot 24, rue Lhomond,
75231 Paris Cedex 05, France }

\author{E. Flurin}

\affiliation{Laboratoire Pierre Aigrain, Ecole Normale Supérieure, CNRS (UMR 8551),
Université P. et M. Curie, Université D. Diderot 24, rue Lhomond,
75231 Paris Cedex 05, France }

\author{A. Auff\` eves}

\affiliation{CNRS and Université Grenoble Alpes, Institut Néel, 38042 Grenoble,
France }

\author{F. Mallet}

\affiliation{Laboratoire Pierre Aigrain, Ecole Normale Supérieure, CNRS (UMR 8551),
Université P. et M. Curie, Université D. Diderot 24, rue Lhomond,
75231 Paris Cedex 05, France }

\author{B. Huard}

\affiliation{Laboratoire Pierre Aigrain, Ecole Normale Supérieure, CNRS (UMR 8551),
Université P. et M. Curie, Université D. Diderot 24, rue Lhomond,
75231 Paris Cedex 05, France }

\date{\today}

\begin{abstract}
The fluorescence of a resonantly driven superconducting qubit is measured in the time domain, providing a weak probe of the qubit dynamics. Prior preparation and final, single-shot measurement of the qubit allows to average fluorescence records conditionally on past and future knowledge. The resulting interferences reveal purely quantum features characteristic of weak values. We demonstrate conditional averages that go beyond classical boundaries and probe directly the jump operator associated with relaxation. The experimental results are remarkably captured by a recent theory, which generalizes quantum mechanics to open quantum systems whose past and future are known.

\end{abstract}

\maketitle

In quantum physics, measurement results are random but their statistics can be predicted assuming some knowledge about the system in the past.  Additional knowledge from a future measurement~\cite{Aharonov1964} deeply changes the statistics in the present and leads to purely quantum features~\cite{Aharonov1988,Aharonov2010}. In particular conditioned average outcomes of a weak measurement, revealing the so-called weak values, were shown to go beyond the classically allowed range and give a way to directly measure complex quantities~\cite{Dressel2013}. Recently, these concepts have been considered in the general case of open quantum systems where decoherence occurs~\cite{Wiseman2002,Tsang2009,Gammelmark2013a}. Then, what are the properties of weak values for the unavoidable measurement associated to decoherence, the one performed by the environment? Here, we answer this question in the simplest open quantum system: a quantum bit in presence of a relaxation channel. We continuously monitor the fluorescence emitted by a superconducting qubit driven at resonance. Conditioned on initial preparation and final single shot measurement outcome of the qubit state, we probe weak values displaying non-classical properties. The fluorescence signal exhibits interferences between oscillations associated to past and future quantum states~\cite{Wiseman2002,Tsang2009,Gammelmark2013a}. The measured data are in complete agreement with theory.

\begin{figure*}
\includegraphics[width=2\columnwidth]{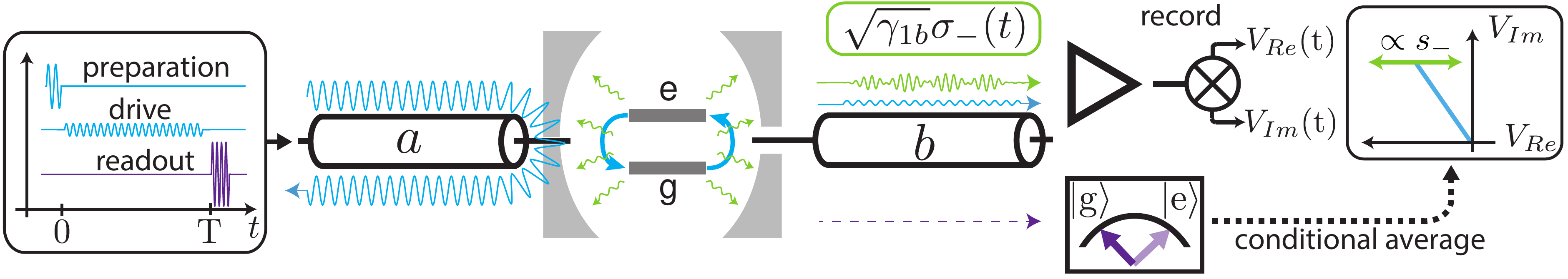}
\caption{\textbf{Principle of the experiment.} A resonant field (blue on line $a$) drives a qubit via a weakly coupled line $a$. While the off-resonant cavity reflects most of the driving field back on line $a$, the fluorescence signal (green) mostly exits through the strongly coupled line $b$ with an amplitude proportional to $\sigma_{-}$, and oscillating at the Rabi frequency $\nu_{R}$. Due to nonzero transmission from line $a$, it is displaced by a resonant field independent of the qubit state (blue on line $b$). This signal is then measured at time $t$ with a heterodyne detection setup including a phase-preserving quantum limited amplifier (triangle). At final time $T$, the qubit state is measured with high fidelity using a pulse at the bare cavity frequency (purple), enabling conditional averaging of the fluorescence signal depending on the measured state. In the quadrature phase space rotating at $\nu_q$ (right panel), resonance fluorescence is revealed by the time oscillation (green) of the voltage $V_\mathrm{Re}$, shifted by a constant value (blue). \label{fig1}}
\end{figure*}

A two-level system irradiated at resonance undergoes Rabi oscillations between ground state $|g\rangle$ and excited state $|e\rangle$. Conversely, these oscillations leave a footprint in the emitted fluorescence field. In the spectral domain, two side peaks appear around resonance frequency, constituting the Mollow triplet~\cite{Mollow1969}. They were first observed in quantum optics and more recently in the microwave range~\cite{Astafiev2010a}. If the detection setup allows monitoring fluorescence in the time domain, one gets a weak probe of the qubit. To access weak values of the associated qubit operator, one additionally needs to post-select the experiments depending on qubit state, which therefore needs to be measured in a single-shot manner. Superconducting qubits in cavity are fit for this task~\cite{Houck2007,Abdumalikov2011,Palacios-Laloy2010,Groen2013}. The principle of our experiment is described in Fig.~\ref{fig1}. A transmon qubit with frequency $\nu_{q}=5.19\ \mathrm{GHz}$ is enclosed in a non-resonant superconducting 3D cavity~\cite{Paik2011}, connected to two transmission lines. Line $a$ is coupled\textcolor{black}{{} as weakly as the internal cavity losses }with a rate $\Gamma_{a}=2~\mathrm{kHz}$. This line is used as a channel for resonant driving of the qubit. Since the fundamental cavity mode is far detuned from the qubit frequency by $\nu_{c}-\nu_{q}=2.57~\mathrm{GHz}$, almost all the resonant incoming signal is reflected. The cavity is coupled more strongly to line $b$, with a rate $\Gamma_{b}=0.25~\mathrm{MHz}$. With such an asymmetric coupling, most of the resonance fluorescence is emitted in the outgoing mode $b_{out}$ and the fluorescence signal is not blinded by the large incoming drive. 

\begin{figure}[htbp!]
\centering{}\includegraphics[width=1\columnwidth]{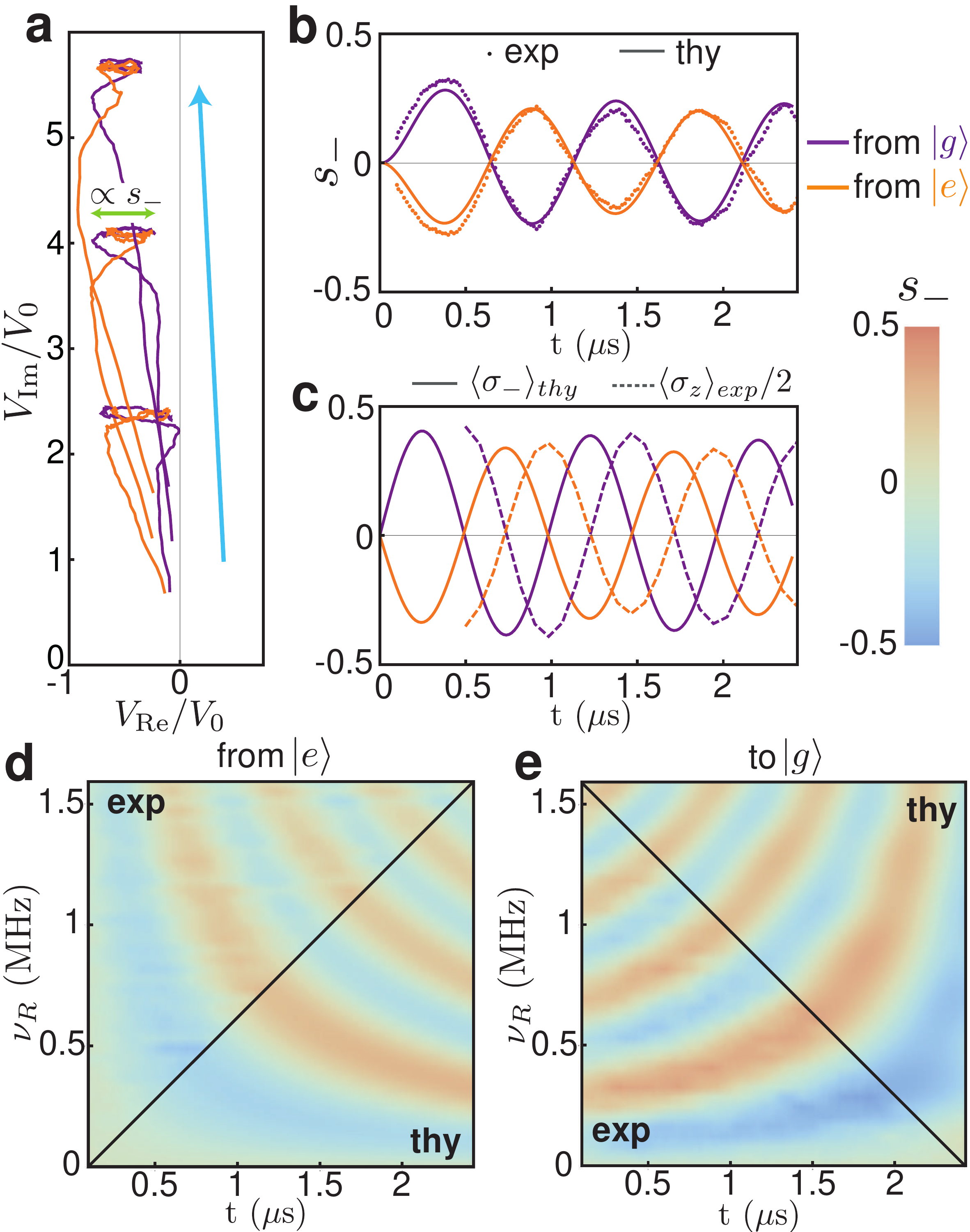}
\caption{\textbf{Resonance fluorescence in time domain. a,} Average time traces of the heterodyned outgoing field on line $b$ represented in the Fresnel plane $(V_\mathrm{Re},V_\mathrm{Im})$, for driving amplitudes corresponding to $\nu_{R}=0.6$, 1 and $1.4\ \mathrm{MHz}$ (blue arrow represents increasing drive amplitudes). The unit of voltage $V_0$ corresponds to an average emitted photon rate equal to $\gamma_{1b}$. Purple (resp. orange) lines correspond to a qubit prepared in $|g\rangle$ (resp. $|e\rangle$) at $t=0$. Each trace is the sum of a term proportional to the drive amplitude and an oscillating part in $V_\mathrm{Re}$ which corresponds to the resonance fluorescence. The finite bandwidth (1.6~MHz) of the detection setup deforms the time traces (finite rise time and diminished oscillation amplitude). \textbf{b,} Dots: Average fluorescence signal $s_-$ as a function of time $t$ for a Rabi frequency $\nu_{R}=1\ \mathrm{MHz}$. Lines: corresponding predicted fluorescence signal filtered by detection setup. \textbf{c,} Dashed lines: measured values of $\langle\sigma_{z}\rangle$ for $\nu_{R}=1\ \mathrm{MHz}$. Plain lines: predicted $\langle\sigma_{-}\rangle$ leading to plain lines in (b). \textbf{d,e,} Average value of the  fluorescence signal $s_-$ as a function of both time and Rabi frequency, for a qubit either prepared in $|e\rangle$ (d) or post-selected in $|g\rangle$ (e). Both measured and predicted averages of $s_-$ are shown in separate regions. Absolute values remain well bellow $0.5$, as expected for the measurement of $\textrm{Re}\left[\left\langle \sigma_{-}\right\rangle \right]=\left\langle \sigma_{x}\right\rangle /2$. Each data point was averaged on at least $3\times10^5$ experiments leading to a maximal standard deviation of 0.05 on $\overline{s_-}$. \label{fig2}} \end{figure}

Fluorescence is due to transitions from excited $|e\rangle$ to ground state $|g\rangle$. The amplitude of the emitted field is then proportional to the average of the lowering operator $\sigma_{-}=|g\rangle\langle e|$ of the qubit. Using the input/output formalism and eliminating the non resonant intracavity field operator, one can show~\cite{Valente2012} that the average field outgoing on line $b$ is given by \begin{equation}
\langle b_{out}\rangle=\langle b_{out}\rangle_{0}-\sqrt{\gamma_{1b}}\langle\sigma_{-}\rangle.\label{eq:bout}
\end{equation}
The first term does not depend on qubit state and oscillates at the resonant drive frequency $\nu_{q}$. In the experiment, it is mostly due to an external parasitic cross-talk (-50~dB) dominating the expected finite transmission of the cavity at frequency $\nu_{q}$. The second term corresponds to the field radiated by the qubit, whose amplitude oscillates at the Rabi frequency $\nu_{R}$ (see Fig.~\ref{fig1}). The prefactor $\gamma_{1b}$ is the spontaneous emission rate into line $b$ set by Purcell effect and is estimated to be of the order of $(50~\mu\mathrm{s})^{-1}$~\cite{Campagne-ibarcq}.

The fluorescence signal was measured using a heterodyne detection setup (see \cite{Campagne-ibarcq}). It records in time two voltage traces $V_\mathrm{Re}(t)$ and $V_\mathrm{Im}(t)$ that are respectively proportional to the quadratures $b_{out}+b_{out}^\dagger$ and $(b_{out}-b_{out}^\dagger)/i$ of the outgoing field on line $b$ at frequency $\nu_{q}$. In Fig.~\ref{fig2}a, the average traces are plotted in the Fresnel plane for three different drive amplitudes, with initial qubit states either in the ground $|g\rangle$ (purple line) or excited state $|e\rangle$ (orange line). As expected from Eq.~(\ref{eq:bout}), the measured amplitude is the sum of a time-independent offset proportional to the drive amplitude and of a fluorescence term oscillating at the Rabi frequency. With our choice of phase reference, $\langle\sigma_{-}(t)\rangle$ is a real number so that it oscillates along the real quadrature $V_\mathrm{Re}$ only. The fluorescence signal $s_-(t)\propto V_\mathrm{Re}(t)-V_\mathrm{Re}^{0}$ can now be defined as the oscillating part of the real quadrature (Fig.~\ref{fig2}b). A single proportionality factor is fixed for the whole set of measurements so that the average $\overline{s_-}(t)$ matches in amplitude the predicted value of $\mathrm{Re}\langle\sigma_{-}(t)\rangle$, which is here simply equal to $\langle\sigma_{-}(t)\rangle$. Note that the finite bandwidth $1.6~\mathrm{MHz}$ of the phase-preserving amplifier needs to be taken into account when calculating the fluorescence signal from the predicted time trace of $\langle\sigma_{-}\rangle$ (plain lines in Fig.~\ref{fig2}c), resulting in a temporally deformed version of the theory, which matches well the measured $\overline{s_-}$ (Fig.~\ref{fig2}b). In addition to the measurement of the fluorescence signal, a complementary probing of the qubit dynamics can be realized by the measurement of the qubit population $\langle\sigma_z\rangle$. The corresponding time trace taken in a separate measurement is shown in Fig.\ref{fig2}c. It is obtained using the high-power readout technique~\cite{Reed2010,Bishop2010}, which uses a final microwave tone at the bare cavity resonance frequency. As expected, initial preparation in ground $|g\rangle$ (purple line) or excited state $|e\rangle$ (orange line) lead to opposite modulations at the Rabi frequency $\nu_{R}=1\ \mathrm{MHz}$. Note that the reduced contrast of the oscillations is due to a finite thermal population of the qubit, leading to $p_0=15.4\%$ in state $\left|g\right\rangle $ when preparing state $|e\rangle$ \cite{Campagne-ibarcq}. Thus, two non-commuting qubit operators can be probed using the fluorescence signal and the conventional qubit population measurement. While the latter is a single-shot, discrete measurement, $s_-$ is a weak, continuous measurement whose strength can be characterized~\cite{Clerk2008} by the measurement rate $\gamma_{1b}$, which is of the order of $0.1~\%$ of the detector bandwidth.

According to Eq.~(\ref{eq:bout}), the observed fluorescence traces can be predicted by calculating the real part of the average value $\mathrm{Tr}\left[\rho(t)\sigma_{-}\right]$ of the lowering operator, where $\rho$ is the density operator of the qubit. Its evolution can be predicted from the preparation $\rho(0)=(1-p_0)|e\rangle\langle e|+p_0|g\rangle\langle g|$ and using the master equation in the Lindblad form~\cite{Lindblad1976} 
\begin{equation}
\frac{d\rho}{dt}=-\frac{i}{\hbar}\left[H,\rho\right]+\gamma_{1}\left(\sigma_{-}\rho\sigma_{+}-\frac{1}{2}\left[\sigma_{+}\sigma_{-}\rho+\rho\sigma_{+}\sigma_{-}\right]\right).\label{Lindblad}
\end{equation} The first term describes the Hamiltonian evolution of the qubit in presence of a drive, with $H=h\nu_{q}\sigma_{z}/2+h\nu_{R}\sigma_{y}/2$ in the rotating frame. We use the standard Pauli operators $\sigma_{z}=|e\rangle\langle e|-|g\rangle\langle g|$, $\sigma_{x}=\left(\sigma_{-}+\sigma_{+}\right)$ and $\sigma_{y}=i\left(\sigma_{-}-\sigma_{+}\right)$. The second term takes into account relaxation with a rate $\gamma_{1}=(16~\mu\mathrm{s})^{-1}$, part of which is due to the spontaneous emission rate $\gamma_{1b}$ introduced in Eq.~(\ref{eq:bout}). The excellent agreement between these predictions (lower half) and data (upper half) is shown in Fig.~\ref{fig2}d, where the average fluorescence signal $\overline{s_-}$ is represented as a function of both time and Rabi frequency, for a qubit prepared at time $0$ close to the excited state. The Rabi oscillations of the qubit are apparent both in time and drive amplitude. Here, relaxation only leads to a slight fading of the oscillation contrast since the duration $T=2.5~\mu\mathrm{s}$ of the experiment is much smaller than $\gamma_1^{-1}$.

Figure \ref{fig2}d represents the fluorescence signal averaged on a large set of experiments with identical initial state at time 0. Dually, one can perform the averaging on all experiments where the qubit is measured at time $T$ in an identical final state, given by the outcome of $\sigma_Z$. One can ensure that there is no prior knowledge by preparing the qubit in the maximally entropic state, half experiments starting with the qubit in the ground state and half in the excited state. Such an averaging conditioned on the future only is shown in Fig.~\ref{fig2}e for a qubit post-selected in the ground state at time $T$. Clearly, Fig.~\ref{fig2}e is the time-reversed of Fig.~\ref{fig2}d which reflects the duality between preparation and postselection.

The final measurement outcome used as a post-selection criterion can be modeled by a positive operator valued measure $E(T)$~\cite{Haroche2006}. For instance, when the measurement of $\sigma_{z}$ indicates that the qubit is in the ground state, $E(T)=(1-p_T)|g\rangle\langle g|+p_T|e\rangle\langle e|$, where $p_T\ll 1$ takes into account the imperfection of the measurement. The post-selected average value of the lowering operator is then given at any time $t$ before $T$ by $\mathrm{Tr}\left[E(t)\sigma_{-}\right]/\mathrm{Tr}\left[E(t)\right]$~\cite{Wiseman2002,Tsang2009,Gammelmark2013a}. Here, we have used a time dependent post-selection operator  $E(t)$, which obeys a similar equation to Eq.~(\ref{Lindblad}) valid for times $t\leq T$
\begin{equation}
\frac{dE}{dt}=-\frac{i}{\hbar}\left[H,E\right]-\gamma_{1}\left(\sigma_{+}E\sigma_{-}-\frac{1}{2}\left[\sigma_{+}\sigma_{-}E+E\sigma_{+}\sigma_{-}\right]\right).
\end{equation}
The corresponding prediction for the post-selected average value of $s_-$ is in excellent agreement with the measured one as shown in Fig.~\ref{fig2}e. Note that the slightly better contrast of the post-selected oscillations compared to the preselected ones is explained by a more efficient measurement than preparation $(p_0>p_T)$.

\begin{figure}[htbp]
\centering{}\includegraphics[width=1\columnwidth]{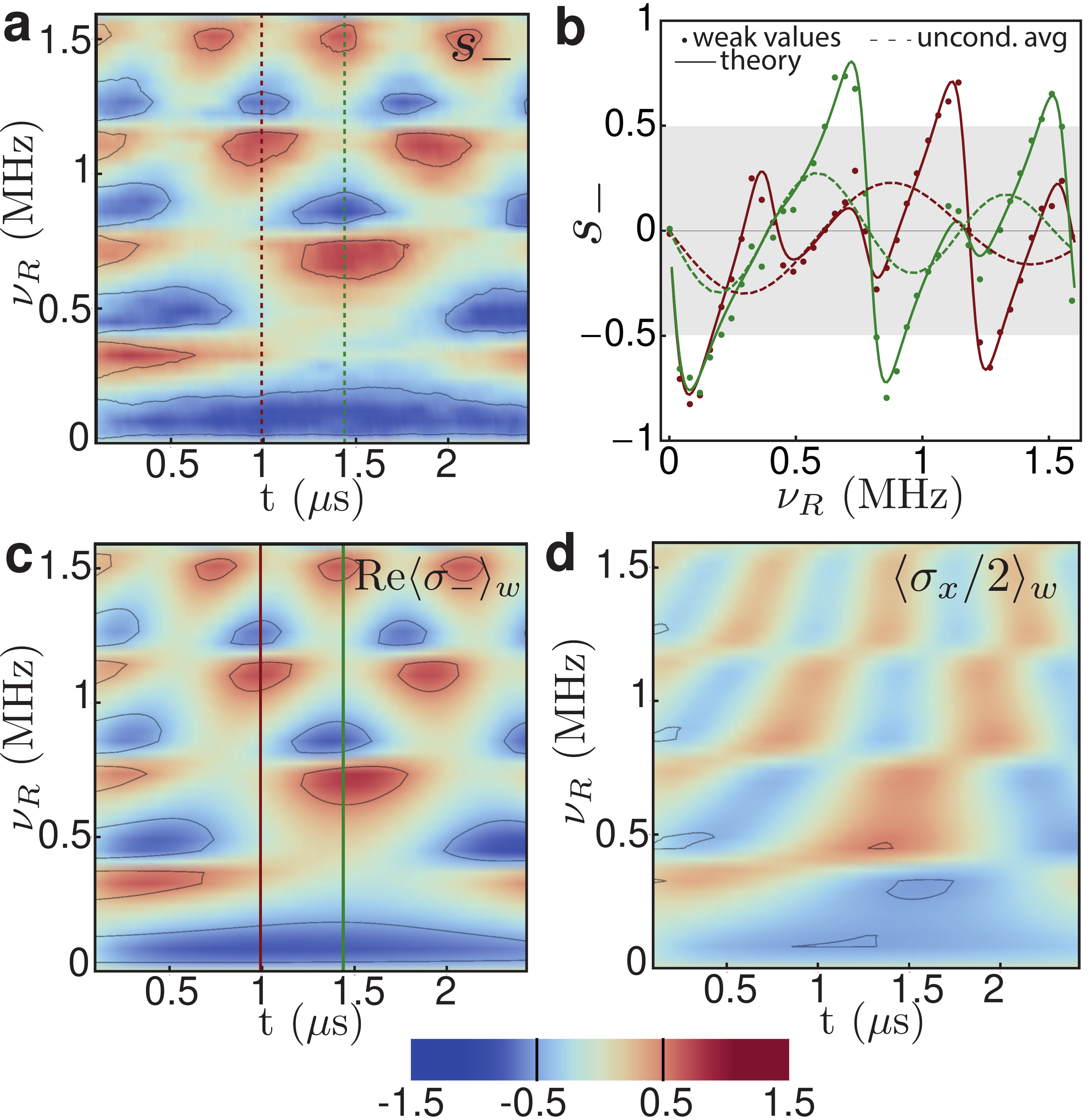}
\caption{\textbf{Interferences between past and future states.} \textbf{a,}  Average value of the measured fluorescence signal $s_-$ as a function of both time and Rabi frequency, for a qubit prepared in $|e\rangle$ and post-selected in $|g\rangle$. Plain lines surround regions with weak values beyond the range allowed by macro realism. \textbf{b,} Dots: cuts of \textbf{a} as a function of $\nu_R$ for times $t=0.99~\mu\mathrm{s}$ (green) and $t=1.44~\mu\mathrm{s}$ (red). The maximal standard deviation on each average of $s_-$ is 0.05. Plain lines: prediction for the same curves using Eq~(\ref{weakvalueeq}). Dashed lines: cuts of Fig.~\ref{fig2}d at the same times. The gray region delimits the range of possible unconditional average values, like the contours in (a).  \textbf{c,} Theoretical counterpart of \textbf{a} assuming that the average of $s_-$ is a measure of $\mathrm{Re}(\langle \sigma_-\rangle_w)$ and using Eq.~\ref{weakvalueeq}. \textbf{d,} Theoretical counterpart of \textbf{a} assuming that the average of $s_-$ is a measure of $\langle\mathrm{Re}\sigma_-\rangle_w)$.}\label{fig3}
\end{figure}

How are time traces of fluorescence modified when using knowledge of both past and future? The conditional average of the fluorescence signal is represented in Fig.~\ref{fig3} for both a preparation in excited state (as in Fig.~\ref{fig2}d) and a postselection in ground state (as in Fig.~\ref{fig2}e). This fluorescence signal, which probes the weak values $\langle\sigma_-\rangle_w$, is dramatically changed. Schematically, Fig.~\ref{fig3}a exhibits interferences between the oscillations of Fig.~\ref{fig2}d and of Fig.~\ref{fig2}e, with the appearance of negative (blue) and positive (red) pockets. There are times $t$ and Rabi frequencies $\nu_R$ in these pockets for which the weak values go beyond the conventional range of unconditional averages, set by $\left|\mathrm{Re}({\langle\sigma_{-}\rangle})\right|\leq1/2$. In Fig.~\ref{fig3}, plain lines represent the contours within which this boundary is violated. Quantitatively, the largest weak value we could obtain is $1.15\pm0.05$ which is well beyond 0.5. This purely quantum effect, first predicted in 1988~\cite{Aharonov1988} and observed already in quantum optics in 1991~\cite{Ritchie1991}, is a complementary evidence to the irrelevancy of macro realism~\cite{Williams2008}. In superconducting circuits, out of bound weak values have already been demonstrated in connection with the Leggett-Garg inequalities on the autocorrelation spectrum of $\sigma_z(t)$~\cite{Palacios-Laloy2010} and for discrete weak measurements performed by another artificial atom~\cite{Groen2013}. 

Special features develop when past and future information disagree, which is for Rabi frequencies such that the qubit rotates by an even amount of $\pi$ in a time $T$ (Fig.~\ref{fig3}). There, the weak values go to zero but a small shift in Rabi frequency results in a dramatic change of the signal as evidenced in Fig.~\ref{fig3}b, where the conditioned average of the fluorescence signal is shown as dots as a function of $\nu_R$ at times $t=0.99~\mu\mathrm{s}$ and $t=1.44~\mu\mathrm{s}$. At the sign change, the slope of the weak value is much stiffer than the one of the unconditional signal (dashed line), which is characteristic of the amplifying abilities of weak values \cite{Hosten2008,Dixon2009}. This curve was the most sensitive way to determine the measurement fidelity of the qubit population at time $T$ in our experiment~\cite{Campagne-ibarcq}. Note that it does not mean that other post-processing techniques than the conditional average would not result in an even better parameter estimation~\cite{Knee2013,Ferrie2014,Tanaka2013,Knee2013a}.

The conditional average of fluorescence signals can be quantitatively understood using the same formalism as described above. The weak value for $\sigma_{-}$ at any time $t$ can indeed be obtained from the operators $\rho(t)$ from the past and $E(t)$ from the future, and is given by~\cite{Wiseman2002,Tsang2009,Gammelmark2013a}. \begin{equation}
\langle\sigma_{-}\rangle_{w}=\mathrm{Tr}(\tilde{\rho}\sigma_{-})\textrm{, where }\tilde{\rho}(t)=\frac{\rho(t) E(t)}{\mathrm{Tr}(\rho(t) E(t))}.\label{weakvalueeq}
\end{equation}
The experiment offers a quantitative test of this simple expression, since the post-selected fluorescence signal is given by $\mathrm{Re}(\langle\sigma_{-}\rangle_{w})$. As can be seen on Fig.~\ref{fig3}b, the resulting prediction (plain lines) agrees well with the data (dots). The agreement is good for all measurements as can be seen between Figs.~\ref{fig3}a and \ref{fig3}c where both prediction and measurements are compared as a function of time $t$ and Rabi frequency $\nu_R$. The predicted contours surrounding the regions where macro realism is violated are represented as plain lines and they indeed match well their experimental counterpart. The agreement was excellent for any conditions we considered on preparation and post-selection~\cite{Campagne-ibarcq}.

Interestingly, the operator $\sigma_-$ probed by the conditional averaged $s_-$ is not an observable as it is not hermitian. This illustrates the ability of conditional averages of weak measurements to probe complex quantities~\cite{Lundeen2011}. Here, the measured observable leading to $s_-$ is the field quadrature $\mathrm{Re}(b_{out})=(b^\dagger_{out}+b_{out})/2$. For averages with either pre-selection or post-selection only, Eq.~(\ref{eq:bout}) leads to $\overline{s_-}=\mathrm{Re}\langle\sigma_-\rangle$ or $\overline{s_-}=\langle\mathrm{Re}\sigma_-\rangle=\langle\sigma_x/2\rangle$, which are formally identical. This is not the case anymore for pre and post-selected measurements for which $\mathrm{Re}\langle\sigma_-\rangle_w$  and $\langle\sigma_x/2\rangle_w$ differ and indeed give very different predictions as can be seen in Figs~\ref{fig3}c and \ref{fig3}d. It is clear that the experiment matches only the prediction associated with $\mathrm{Re}\langle\sigma_-\rangle_w$, which cannot be interpreted as the weak value of the observable $\sigma_x/2$.

In conclusion, we demonstrated that detecting resonance fluorescence radiated by a superconducting qubit out of a cavity corresponds to a weak continuous monitoring of the $\sigma_{-}$ operator of the qubit. Using conditional averaging on the fluorescence signal depending on the measured final state of the qubit, we observed interferences between Rabi oscillations associated to past and future states. The experiment offers a quantitative demonstration of the accuracy of recent theoretical works~\cite{Wiseman2002,Tsang2009,Gammelmark2013a} able to predict the conditional average of continuous recording in open quantum systems. Fluorescence tracking illustrates several key aspects of weak values: violation of macro realism, improvement of parameter estimation and non-hermitian operator measurement. Besides, by recording efficiently the fluorescence signal, one should be able to fully estimate the qubit trajectory. It may be a way to correct for relaxation in real time by feedback~\cite{Vijay2012c,Riste2012,Campagne-Ibarcq2013,deLange2014} as long as decoherence is limited by emission into a transmission line.

\begin{acknowledgements}
We thank Michel Devoret, Vladimir Manucharyan, Mazyar Mirrahimi, Pierre
Rouchon, Benoit Douçot, Klaus M\o lmer, Brian Julsgaard, Pascal Degiovanni, Daniel Valente and Patrice
Bertet for enlightening discussions. Nanofabrication has been made
within the consortium Salle Blanche Paris Centre. This work was supported
by the ANR contracts ANR-12-JCJC-TIQS and ANR-13-JCJC-INCAL. LB acknowledges support from
Direction Générale de l'Armement.
\end{acknowledgements}

%


\medskip{}
\end{document}